\documentclass{JAC2003}

\usepackage{graphicx}
\usepackage{booktabs}
\usepackage{multicol}
\usepackage{amsmath}
\usepackage{url}
\begin{document}

\title{Review of the multilayer coating model}

\author{Takayuki Kubo\thanks{kubotaka@post.kek.jp}, Takayuki Saeki,  
KEK, High Energy Accelerator Research Organization, \\ Tsukuba, Ibaraki 305-0801, Japan\\
Yoshihisa Iwashita, Institute for Chemical Research, Kyoto University, Uji, Kyoto 611-0011, Japan}

\maketitle

\begin{abstract}
The recent theoretical study on the multilayer-coating model published in Applied Physics Letters~\cite{Kubo} is reviewed. 
Magnetic-field attenuation behavior in a multilayer coating model is different from a semi-infinite superconductor and a superconducting thin film. 
This difference causes that of the vortex-penetration field at which the Bean-Livingston surface barrier disappears.  
A material with smaller penetration depth, such as a pure Nb, is preferable as the substrate for pushing up  the vortex-penetration field of the superconductor layer. 
The field limit of the whole structure of the multilayer coating model is limited not only by the vortex-penetration field of the superconductor layer, but also by that of the substrate. 
Appropriate thicknesses of layers can be extracted from contour plots of the field limit of the multilayer coating model given in Ref.~\cite{Kubo}. 
\end{abstract}

%%%%%%%%%%%%%%%%%%%%%%
%%%%%%%%%%%%%%%%%%%%%%
\section{Introduction}\label{section:introduction}
%%%%%%%%%%%%%%%%%%%%%%
%%%%%%%%%%%%%%%%%%%%%%

The multilayer coating is one of approaches for pushing up the field limit of superconducting (SC) accelerating cavity~\cite{Gurevich}, 
which consists of alternating layers of SC layers ($\mathcal{S}$) and insulator layers ($\mathcal{I}$) formed on a bulk-SC substrate. 
A theoretical understanding on this topic showed progresses last year~\cite{Kubo, Kubo_IPAC13, Kubo_SRF2013}. 
The magnetic-field distribution in multilayer SC was derived by solving the Maxwell and the London equations with correct boundary conditions~\cite{Kubo, Kubo_IPAC13};
forces acting on a vortex and resultant vortex-penetration field, at which the Bean-Livingston surface barrier disappears, were evaluated based on the correct magnetic-field distribution~\cite{Kubo, Kubo_SRF2013};  
and then appropriate choices of layer thicknesses and materials to enhance the field limit were revealed~\cite{Kubo}. 
The above results were then reproduced by an other group~\cite{Posen}.

In this paper, we review Ref.~\cite{Kubo}. 
Especially the  Bean-Livingston surface barrier is explained in detail by comparing those of a semi-infinite SC, an SC thin film, and an $\mathcal S$ layer of the multilayer SC.  
Based on the above, some remarks for planning experiments, such as choices of layer thicknesses and a material combination are described. 
The surface resistance of the multilayer SC is also commented.

%%%%%%%%%%%%%%%%%%%%%%%%%%%%%%%%%%%%%%%%%%%%%%%%%%%%%
\section{Surface barrier}
%%%%%%%%%%%%%%%%%%%%%%%%%%%%%%%%%%%%%%%%%%%%%%%%%%%%%

%
\begin{figure}[*t]
   \begin{center}
   \includegraphics[width=0.8\linewidth]{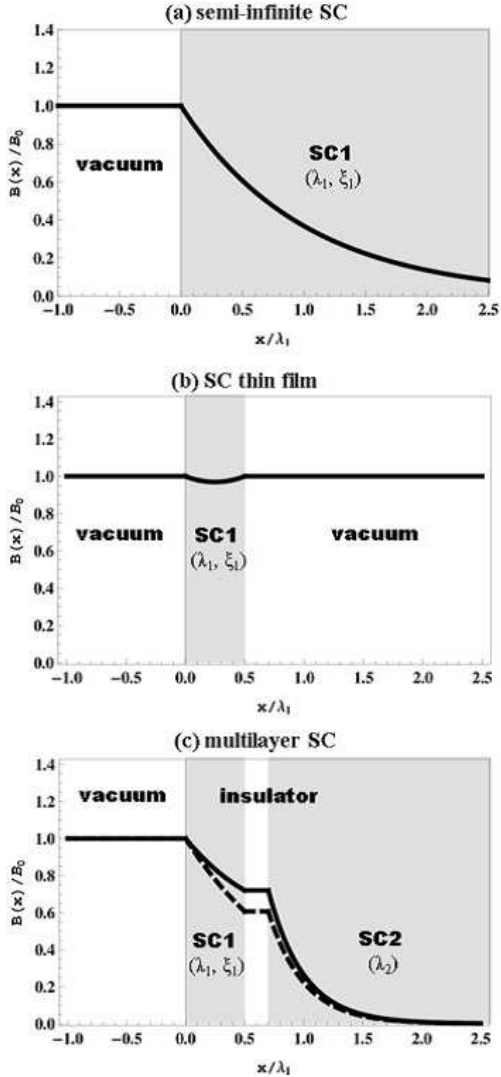}
   \end{center}\vspace{-0.2cm}
   \caption{
Magnetic-fields in three different systems: 
(a) semi-infinite SC, (b) SC thin film, and (c) multilayer SC that consists of a single $\mathcal S$ layer and a single $\mathcal I$ layer.  
The surfaces of materials are assumed to be flat and parallel to the $y$-$z$ plane. 
SC1 is an extreme type II SC material with a penetration depth $\lambda_{1}$ and a coherence length $\xi_1$ ($\xi_1 \ll \lambda_1$) and
SC2 is an arbitrary SC material with a penetration depth $\lambda_{2}$. 
A thickness of the SC thin film and that of the $\mathcal S$ layer of the multilayer SC are assumed to be $d,\, d_{\mathcal S}\gg\xi_1$ and that of the $\mathcal I$ layer is assumed to be $d_{\mathcal I} >$ a few nm. 
Black curves represent magnetic-field distributions. 
A dotted curve in (c) represents an exponential decay for comparison with the correct curve. 
   }\label{fig1}
\end{figure}
%

%%%%%%%%%%%%%%%%%%%%%%%%%%%%%%%%%%%%%%%%%%%%%%%%%%%%%
\subsection{Magnetic Field Distributions} 
%%%%%%%%%%%%%%%%%%%%%%%%%%%%%%%%%%%%%%%%%%%%%%%%%%%%%

By solving the Maxwell and the London equations with appropriate boundary conditions, 
a magnetic-field distribution in a system can be derived. 
Figure~\ref{fig1} shows magnetic-field distributions under the surface magnetic-field ${\bf B}_0=(0,0,B_0)$ for a semi-infinite SC, an SC thin-film and a multilayer SC. 
Their analytical expressions are given by 
\begin{eqnarray}
\!\!\!\!\!\!\!\!\!\!\!\!\!\!\!\!&&B_{\rm [Fig.\ref{fig1}(a)]} 
= B_0 e^{-\frac{x}{\lambda_1}} \, , 
\label{eq:field(a)}  \\
\!\!\!\!\!\!\!\!\!\!\!\!\!\!\!\!&&B_{\rm [Fig.\ref{fig1}(b)]} 
= B_0 \frac{\cosh (\frac{x}{\lambda_1}-\frac{d_{\mathcal S}}{2\lambda_1})}{\cosh\frac{d_{\mathcal S}}{2\lambda_1}} \, , 
\label{eq:field(b)} \\
\!\!\!\!\!\!\!\!\!\!\!\!\!\!\!\!&&B_{\rm [Fig.\ref{fig1}(c)]}^{\mathcal (S)} = B_0
\frac{ \cosh \frac{d_{\mathcal{S}}-x}{\lambda_1} + (\frac{\lambda_2}{\lambda_1} + \frac{d_{\mathcal{I}}}{\lambda_1}) \sinh \frac{d_{\mathcal{S}}-x}{\lambda_1}}
     { \cosh \frac{d_{\mathcal{S}}}{\lambda_1} + (\frac{\lambda_2}{\lambda_1} + \frac{d_{\mathcal{I}}}{\lambda_1}) \sinh \frac{d_{\mathcal{S}}}{\lambda_1}} \, . 
\label{eq:field(c)}
\end{eqnarray}
Note that Eq.~(\ref{eq:field(c)}) represents the magnetic field in the $\mathcal S$ layer ($0\le x \le d_{\mathcal S}$). 
Fields in other rigions ($x>d_{\mathcal S}$) are found in Ref.~\cite{Kubo, Kubo_IPAC13}. 
As shown in Fig.\ref{fig1}(a)-(c) and Eq.~(\ref{eq:field(a)})-(\ref{eq:field(c)}), 
a behavior of magnetic-field attenuation depends on a system, 
which is essential for understanding difference of surface barrier among different systems.

%%%%%%%%%%%%%%%%%%%%%%%%%%%%%%%%%%%%%%%%%%%%%%%%%%%%%
\subsection{Surface Barriers} 
%%%%%%%%%%%%%%%%%%%%%%%%%%%%%%%%%%%%%%%%%%%%%%%%%%%%%

Suppose there exist a vortex with the flux quantum $\phi_0=2.07\times10^{-15}{\rm Wb}$ parallel to ${\bf \hat{z}}$ at a surface of SC. 
This vortex feels two distinct forces ${\bf f}_{\rm I}$ and ${\bf f}_{\rm M}$, 
where ${\bf f}_{\rm I}$ is a force from an image current due to an image antivortex, and
${\bf f}_{\rm M}$ is that from a Meissner current due to an applied magnetic-field.  
The force  ${\bf f}_{\rm I}$ is common in all configurations of Fig.~\ref{fig1} if $\xi_1 \ll d_{\mathcal S}$. 
The derivation is reviewed in detail in Ref.~\cite{Kubo_SRF2013}, 
in which ${\bf f}_{\rm I}$ is given by 
\begin{eqnarray}
{\bf f}_{\rm I} = - \frac{\phi_0^2}{4\pi\mu_0 \lambda_1^2 \xi_1} {\bf \hat{x}} \, , \label{eq:imageforce} \nonumber
\end{eqnarray}
where $\mu_0=4\pi\times10^{-7}{\rm H/m}$ is the magnetic permeability of vacuum. 
Thus the image antivortex attracts the vortex to the surface and prevents the vortex penetration.  
The other force, ${\bf f}_{\bf M}$, is obtained by evaluating the product ${\bf f}_{\bf M} = {\bf J}_{\rm M}\times \phi_0 {\bf \hat{z}}$, 
where ${\bf J}_{\rm M}=(0,\,-\mu_0^{-1} dB/dx ,\, 0)$ is a Meissner-current density at the vortex position $x\simeq 0$. 
By using Eq.~(\ref{eq:field(a)}), (\ref{eq:field(b)}) and (\ref{eq:field(c)}), we find
\begin{eqnarray}
\!\!\!\!\!\!\!\!\!\!\!\!&&{\bf f}_{\rm M \, [{\rm Fig.\ref{fig1}(a)}]}
\!\!=\!\!  \frac{B_0\phi_0}{\mu_0\lambda_1} {\bf \hat{x}} \,, 
\nonumber \\
\!\!\!\!\!\!\!\!\!\!\!\!&&{\bf f}_{\rm M \, [{\rm Fig.\ref{fig1}(b)}] }
\!\!=\!\!  \frac{B_0\phi_0}{\mu_0\lambda_1} \tanh\frac{d}{2\lambda_1} {\bf \hat{x}} 
\simeq \frac{B_0\phi_0}{\mu_0\lambda_1}\frac{d_{\mathcal S}}{2\lambda_1} {\bf \hat{x}}  \,, 
\nonumber \\
\!\!\!\!\!\!\!\!\!\!\!\!&&{\bf f}_{\rm M\, [{\rm Fig.\ref{fig1}(c)}] }^{\mathcal (S)}
\!\!=\!\! \frac{B_0\phi_0}{\mu_0 \lambda_1}
\frac{ \lambda_1 \sinh \frac{d_{\mathcal{S}}-x}{\lambda_1} \!+\! (\lambda_2 \!+\! d_{\mathcal{I}}) \cosh \frac{d_{\mathcal{S}}-x}{\lambda_1}}
{\lambda_1 \cosh \frac{d_{\mathcal{S}}}{\lambda_1} \!+\! (\lambda_2 \!+\! d_{\mathcal{I}}) \sinh \frac{d_{\mathcal{S}}}{\lambda_1}} {\bf \hat{x}} \, 
\nonumber
\end{eqnarray}
by which the vortex is attracted to the inside of each SC. 
The total force acting on the vortex is given by ${\bf f}_{\rm tot} ={\bf f}_{\rm I}+{\bf f}_{\rm M}$. 
When $B_0$ is so small that $|{\bf f}_{\rm M}| < |{\bf f}_{\rm I}|$, 
the force ${\bf f}_{\rm tot}$ directs the negative direction of the $x$-axis, 
which acts as a barrier that prevents the vortex penetration (Bean-Livingston surface barrier). 
When $B_0$ is so large that $|{\bf f}_{\rm M}| > |{\bf f}_{\rm I}|$, 
the barrier disappears and the vortex is drawn into SC. 
Then $B_v$, the surface magnetic-field where the Bean-Livingston barrier disappears, can be obtained by balancing the two forces: 
\begin{eqnarray}
\!\!\!\!\!\!\!\!\!\!\!\!\!\!\!\!\!\!\!\!\!&&
B_{v\,{\rm [Fig.\ref{fig1}(a)]} }
= \frac{\phi_0}{4\pi \lambda_1 \xi_1}  \,\Bigl( \equiv B_{v0} \Bigr) 
\,,  \label{eq:BvBulk}  \\
\!\!\!\!\!\!\!\!\!\!\!\!\!\!\!\!\!\!\!\!\!&&
B_{v\,{\rm [Fig.\ref{fig1}(b)] } }
= \frac{\phi_0}{2\pi d_{\mathcal S}\, \xi_1}
= \frac{2\lambda_1}{d_{\mathcal S}} B_{v0} \,, \label{eq:BvFilm}  \\
\!\!\!\!\!\!\!\!\!\!\!\!\!\!\!\!\!\!\!\!\!&&
B_{v\,{\rm [Fig.\ref{fig1}(c)] }}^{ ({\mathcal S} ) }
\!\!=\!\! 
  \frac{\cosh\frac{d_{\mathcal{S}}}{\lambda_1} \!+\! (\frac{\lambda_2}{\lambda_1} \!+\! \frac{d_{\mathcal{I}}}{\lambda_1})\sinh\frac{d_{\mathcal{S}}}{\lambda_1}}
       {\sinh\frac{d_{\mathcal{S}}}{\lambda_1} \!+\! (\frac{\lambda_2}{\lambda_1} \!+\! \frac{d_{\mathcal{I}}}{\lambda_1})\cosh\frac{d_{\mathcal{S}}}{\lambda_1} }
B_{v0} 
\equiv B_{v}^{ ({\mathcal S} ) } \!.
\label{eq:BvML}
\end{eqnarray}
Eq.~(\ref{eq:BvBulk}) is the well-known result for the semi-infinite SC~\cite{BeanLivingston, GurevichCiovati}, 
Eq.~(\ref{eq:BvFilm}) corresponds to the result shown in Ref.~\cite{Stejic}, and 
Eq.~(\ref{eq:BvML}) is the vortex-penetration field of the top $\mathcal{S}$ layer of the multilayer SC~\cite{Kubo, Kubo_SRF2013}.

Differences among Eq.~(\ref{eq:BvBulk}), (\ref{eq:BvFilm}) and (\ref{eq:BvML}) are due to those of slopes of magnetic-field attenuation at the surfaces, because the force pushing a vortex into SC is given by $|{\bf f}_{\rm M}| \propto |{\bf J}_{\rm M}| \propto |dB/dx|$. 
A smaller $|dB/dx|_{x=0}$ induces a larger $B_v$. 
In fact $|dB/dx|_{x=0}$ of the SC thin film is smaller than that of the semi-infinite SC as shown in Fig.~\ref{fig1}(a) and (b), and Eq.~(\ref{eq:BvFilm}) is larger than Eq.~(\ref{eq:BvBulk}) by a factor $2\lambda_1/d_{\mathcal S}$. 
Similarly, when $|dB/dx|_{x=0}$ of the $\mathcal S$ layer of multilayer SC is smaller than that of the semi-infinite SC, 
Eq.(\ref{eq:BvML}) can be larger than Eq.~(\ref{eq:BvBulk}).

%%%%%%%%%%%%%%%%%%%%%%%%%%%%%%%%%%%%%%%%%%%%%%%%%%%%%%%%%%%%%%%%%
%%%%%%%%%%%%%%%%%%%%%%%%%%%%%%%%%%%%%%%%%%%%%%%%%%%%%%%%%%%%%%%%%
\section{Toward experiments} 
%%%%%%%%%%%%%%%%%%%%%%%%%%%%%%%%%%%%%%%%%%%%%%%%%%%%%%%%%%%%%%%%%
%%%%%%%%%%%%%%%%%%%%%%%%%%%%%%%%%%%%%%%%%%%%%%%%%%%%%%%%%%%%%%%%%

%%%%%%%%%%%%%%%%%%%%%%%%%%%%%%%%%%%%%%%%%%%%%%%%%%%%%%%%%%%%%%%%%
%%%%%%%%%%%%%%%%%%%%%%%%%%%%%%%%%%%%%%%%%%%%%%%%%%%%%%%%%%%%%%%%%
\subsection{Surface Barrier of the $\mathcal S$ Layer} 
%%%%%%%%%%%%%%%%%%%%%%%%%%%%%%%%%%%%%%%%%%%%%%%%%%%%%%%%%%%%%%%%%
%%%%%%%%%%%%%%%%%%%%%%%%%%%%%%%%%%%%%%%%%%%%%%%%%%%%%%%%%%%%%%%%%

\begin{figure}[t]
   \begin{center}
   \includegraphics[width=1\linewidth]{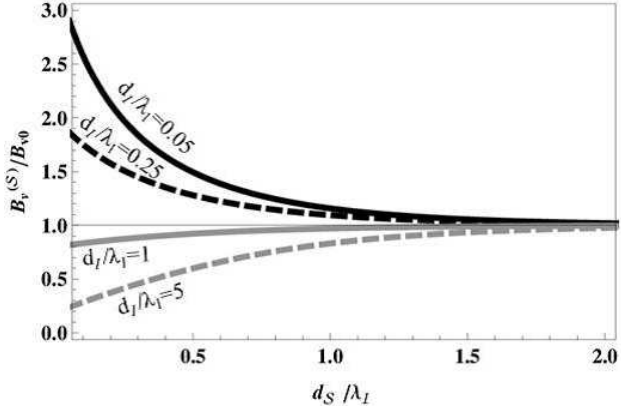}
   \end{center}\vspace{-0.2cm}
   \caption{
Enhancement factors $B_v^{\mathcal (S)}/B_{v0}$ as functions of $d_{\mathcal S}/\lambda_1$,  
where a penetration depth of the bulk-SC substrate is assumed to be $\lambda_2 =0.2 \lambda_1$. 
   }\label{fig2}
\end{figure}

Figure~\ref{fig2} shows enhancement factor $B_v^{\mathcal (S)}/B_{v0}$ as functions of $d_{\mathcal S}/\lambda_1$. 
A combination of small $d_{\mathcal S}/\lambda_1$ and $d_{\mathcal I}/\lambda_1$ yields a large enhancement. 
Substituting $d_{\mathcal S}/\lambda_1 \ll1$ and $d_{\mathcal I}/\lambda_1\ll 1$ into Eq.~(\ref{eq:BvML}), 
we find
\begin{eqnarray}
B_{v}^{ ({\mathcal S} ) } \Bigr|_{\frac{d_{\mathcal S}}{\lambda_1},\frac{d_{\mathcal I}}{\lambda_1}\ll1}
\simeq \biggl( \frac{\lambda_1}{\lambda_2} \biggr) B_{v0}\, . \label{eq:BvMLmax}
\end{eqnarray}
Eq.~(\ref{eq:BvMLmax}) tells the importance of a choice of bulk-SC substrate~\footnote{Note that Eq.~(\ref{eq:BvMLmax}) ceases to be valid at $d_{\mathcal S} \sim \xi_1$ and $d_{\mathcal I} \sim$ a few nm, at which the model should be reevaluated by more accurate theories. }: 
a material with smaller $\lambda_2$, such as a pure Nb with a long mean free path, should be chosen for an enhancement of $B_{v}^{ ({\mathcal S} ) }$.

%%%%%%%%%%%%%%%%%%%%%%%%%%%%%%%%%%%%%%%%%%%%%%%%%%%%%%%%%%%%%%%%%
%%%%%%%%%%%%%%%%%%%%%%%%%%%%%%%%%%%%%%%%%%%%%%%%%%%%%%%%%%%%%%%%%
\subsection{Field Limit of Multilayer SC}
%%%%%%%%%%%%%%%%%%%%%%%%%%%%%%%%%%%%%%%%%%%%%%%%%%%%%%%%%%%%%%%%%
%%%%%%%%%%%%%%%%%%%%%%%%%%%%%%%%%%%%%%%%%%%%%%%%%%%%%%%%%%%%%%%%%

%
\begin{figure}[t]
   \begin{center}
  \includegraphics[width=0.95\linewidth]{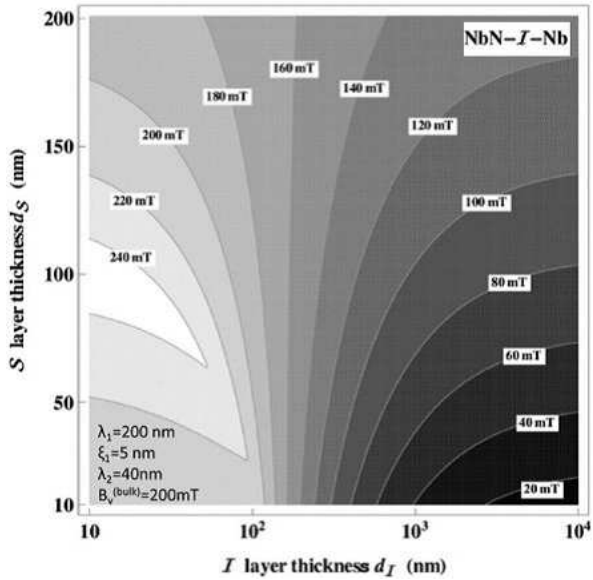}
   \end{center}\vspace{-0.2cm}
   \caption{
An example of contour plots of $B_v^{\rm (ML)}$~\cite{Kubo}. 
Material parameters are assumed to be $\lambda_1=200\,{\rm nm}$ and $\xi_1=5\,{\rm nm}$ for NbN layer. 
The bulk-SC substrate is assumed to be Nb with $\lambda_2=40\,{\rm nm}$ and $B_{v}^{ ({\rm bulk})}=200\,{\rm mT}$.
   }\label{fig3}
\end{figure}

The field limit of the whole structure of the multilayer SC, $B_v^{\rm (ML)}$, is limited not only by $B_{v}^{ ({\mathcal S} ) } $, but also by that of the bulk-SC substrate, $B_{v}^{ ({\rm bulk})}$, 
because the magnetic-field is not completely shielded by the $\mathcal S$ layer alone and that on the interface of the bulk-SC substrate $B_i$ is finite. 
$B_i$ is given by $B_i  = \alpha B_0$, 
where $\alpha = [ \cosh \frac{d_{\mathcal{S}}}{\lambda_1} + (\frac{\lambda_2}{\lambda_1} + \frac{d_{\mathcal{I}}}{\lambda_1}) \sinh \frac{d_{\mathcal{S}}}{\lambda_1} ]^{-1}$. 
If $B_i$ is larger than $B_{v}^{ ({\rm bulk} ) }$, the bulk-SC substrate can also suffer a vortex penetration. 
Thus $B_v^{\rm (ML)}$ is given by
\begin{eqnarray}
B_v^{\rm (ML)} 
\!\!=\!\!
  \begin{cases}
    B_{v}^{ ({\mathcal S} ) }  & (\alpha B_{v}^{ ({\mathcal S} ) } < B_v^{\rm (bulk)}) \\
    \alpha^{-1} B_v^{\rm (bulk)}  & (\alpha B_v^{ ({\mathcal S} ) } \ge B_v^{\rm (bulk)}) . 
  \end{cases}
\end{eqnarray}
Fig.~\ref{fig3} shows examples contour plots of  $B_v^{\rm (ML)}$ given in Ref.~\cite{Kubo}, 
from which an appropriate combination of $\mathcal S$ and $\mathcal I$ layer thicknesses can be found.

%%%%%%%%%%%%%%%%%%%%%%%%%%%%%%%%%%%%%%%%%%%%%%%%%%%%%%%%%%%%%%%%%
%%%%%%%%%%%%%%%%%%%%%%%%%%%%%%%%%%%%%%%%%%%%%%%%%%%%%%%%%%%%%%%%%
\subsection{Surface Resistance}
%%%%%%%%%%%%%%%%%%%%%%%%%%%%%%%%%%%%%%%%%%%%%%%%%%%%%%%%%%%%%%%%%
%%%%%%%%%%%%%%%%%%%%%%%%%%%%%%%%%%%%%%%%%%%%%%%%%%%%%%%%%%%%%%%%%

Not only the field-limit, but also the quality factor is expected to be improved by the multilayer coating,  
because parts of currents flow in an $\mathcal S$ layer material with a small dissipation. 
This effect modifies the surface resistance formula, 
but its derivation is not trivial.  
The formula given in Ref.~\cite{Gurevich} is based on the assumption of exponential decay of the magnetic field, 
which is not necessarily a good approximation and should be reevaluated by using the correct field-distribution and formalism given in Ref.~\cite{Kubo, Kubo_IPAC13}. 
The derived formula and detailed discussions will be presented elsewhere~\cite{Kubo_elsewhere}. 

%%%%%%%%%%%%%%%%%%%%%%%%%%%%%%%%%%%%%%%%%%%%%%%%%%%%%%%%%%%%%%%%%
%%%%%%%%%%%%%%%%%%%%%%%%%%%%%%%%%%%%%%%%%%%%%%%%%%%%%%%%%%%%%%%%%
\section{Summary}
%%%%%%%%%%%%%%%%%%%%%%%%%%%%%%%%%%%%%%%%%%%%%%%%%%%%%%%%%%%%%%%%%
%%%%%%%%%%%%%%%%%%%%%%%%%%%%%%%%%%%%%%%%%%%%%%%%%%%%%%%%%%%%%%%%%

In this paper, we have reviewed the multilayer coating model~\cite{Kubo, Kubo_IPAC13, Kubo_SRF2013}. 
\begin{itemize}
\item Magnetic-field attenuation behavior in a multilayer SC is different from a semi-infinite SC and an SC thin film. This difference causes a difference of the vortex-penetration field at which the Bean-Livingston surface barrier disappears.   
\item A material with smaller penetration depth is preferable as the bulk-SC substrate for pushing up  the vortex-penetration field of the $\mathcal S$ layer, $B_{v}^{ ({\mathcal S} ) } $. 
\item The field limit of the whole structure of the multilayer SC, $B_v^{\rm (ML)}$, is limited not only by $B_{v}^{ ({\mathcal S} ) } $, but also by that of the bulk-SC substrate, $B_{v}^{ ({\rm bulk})}$. Appropriate thicknesses of $\mathcal S$ and $\mathcal I$ layers can be extracted from contour plots of $B_v^{\rm (ML)}$ given in Ref.~\cite{Kubo}. 
\end{itemize}


\begin{thebibliography}{99}
   
\bibitem{Kubo}
T. Kubo, Y. Iwashita, and T. Saeki,  
Appl. Phys. Lett. {\bf 104}, 032603 (2014).

\bibitem{Kubo_IPAC13}
T. Kubo, Y. Iwashita, and T. Saeki, 
Rf Field-Attenuation Formulae for the Multilayer Coating Model, 
in {\it Proceedings of IPAC'13, Shanghai, China} (2013), p. 2343, WEPWO014.   

\bibitem{Kubo_SRF2013}
T. Kubo, Y. Iwashita, and T. Saeki, 
Vortex Penetration Field in the Multilayered Coating Model, 
in {\it Proceedings of SRF2013, Paris, France} (2013), p. 427, TUP007. 

\bibitem{Gurevich}
A. Gurevich, 
Appl. Phys. Lett. {\bf 88}, 012511 (2006).

\bibitem{Posen}
S. Posen et al., 
Theoretical Field Limits for Multi-Layer Superconductors, 
in {\it Proceedings of SRF2013, Paris, France} (2013), p. 788, WEIOC04.   

\bibitem{BeanLivingston}
C. P. Bean and J. D. Livingston, 
Phys. Rev. Lett. {\bf 12}, 14 (1964). 

\bibitem{GurevichCiovati}
A. Gurevich and G. Ciovati, 
Phys. Rev. B {\bf 77}, 104501 (2008). 

\bibitem{Stejic}
G. Stejic, A. Gurevich, E. Kadyrov, D. Christen, R. Joynt, and D.C. Larbalestier, 
Phys. Rev. B {\bf 49}, 1274 (1994). 

\bibitem{Kubo_elsewhere}
T. Kubo, Y. Iwashita, and T. Saeki, to be presented elsewhere.  

\end{thebibliography}
\end{document}